\documentclass[prb,showpacs,twocolumn,amsmath,amssymb]{revtex4}

\usepackage{graphicx}
\usepackage{color}

\usepackage[latin1]{inputenc}
\usepackage{graphicx}
\usepackage{amsmath}
\usepackage{amssymb}
\usepackage{relsize}

\newcommand{\pd}[2]{\frac{\partial #1}{\partial #2}}
\newcommand{\mean}[1]{\left \langle #1 \right \rangle}

\newcommand{\lam}{\lambda}
\newcommand{\beq}{\begin{equation}}
\newcommand{\eeq}{\end{equation}}
\newcommand{\al} {\alpha}
\newcommand{\Wc} {W_{\rm{chem}}}
\newcommand{\cla} {r}
\newcommand{\cra} {s}
\newcommand{\clx} {p}
\newcommand{\crx} {q}
\newcommand{\Nal} {N_\al}
\newcommand{\Nrho} {N_\rho}
\newcommand{\Nj} {N_j}
\newcommand{\Nmu} {N_\mu}
\newcommand{\n} {\mathbf{n}}
\newcommand{\ca} {\mathbf{c}}
\newcommand{\sm} {s_{\rm{m}}}

\newcommand{\stot} {s_{\rm{tot}}}

\newcommand{\ceq} {c^{\rm{eq}}}
\newcommand{\caeq} {\mathbf{c}^{\rm{eq}}}
\newcommand{\eq} {\rm{eq}}
\newcommand{\nn} {\nonumber}
\newcommand{\Qhk} {Q_{\rm{hk}}}
\newcommand{\Qex} {Q_{\rm{ex}}}
\newcommand{\Lv} {\mathbf{L}}
\newcommand{\dbv}{\omega}
\newcommand{\Om}{{\left(\Omega / \dbv_j \right)}}
\newcommand{\calph}{\left( c_\al \dbv_\al\right)}
\newcommand{\Oom}{\left(\Omega / \mathbf{\dbv} \right)}

\begin{document}

%opening
\title{Stochastic thermodynamics of chemical reaction networks}
\author{Tim Schmiedl and Udo Seifert}
\affiliation{{II.} Institut f\"ur Theoretische Physik, Universit\"at Stuttgart,
  70550 Stuttgart, Germany}

\begin{abstract}
 
For chemical reaction networks in a dilute solution described by a master equation, we define energy and entropy on a stochastic trajectory and develop a consistent nonequilibrium thermodynamic description along a single stochastic trajectory of reaction events. A first-law like energy balance relates internal energy, applied (chemical) work and dissipated heat for every single reaction. Entropy production along a single trajectory involves a sum over changes in the entropy of the network itself and the entropy of the medium. The latter is given by the exchanged heat identified through the first law. Total entropy production is constrained by an integral fluctuation theorem for networks arbitrarily driven by time-dependent rates and a detailed fluctuation theorem for networks in the steady state. Further exact relations like a generalized Jarzynski relation and a generalized Clausius inequality are discussed. We illustrate these results for a three-species cyclic reaction network which exhibits nonequilibrium steady states as well as transitions between different steady states.

\end{abstract}
%\today
\maketitle

\section {Introduction}

Networks of coupled chemical reactions in a dilute solution should be described
by a chemical master equation whenever fluctuations are relevant
due to small numbers
of at least one of the involved species \cite{vankampen, gardiner, nicolis,schn76, mou86, gasp04, andr04}. 
As essential parameters
this equation contains the rate constants of all possible reactions.
The solution of the chemical master equation gives 
 the dynamics of
the probability to find  a  certain number of molecules of each species
at a given time for a given initial condition. The stochastic character
of such a description implies that for each realization we can talk
about the stochastic trajectory of the network by recording, in principle,
the time at which each particular reaction took place with its
concomitant change of the number of molecules.  
Even though such 
a description can be applied to  networks in thermodynamic equilibrium,
it is particularly relevant for driven networks in open systems
 where externally imposed boundary conditions
generate fluxes even at a constant (mean) concentration of species.
More generally, such a description holds  true even if the rate ``constants'' 
become time-dependent, e.g., if the concentration of a chemical species
can be controlled externally in a time-dependent fashion.

Quite often, 
such networks are effectively at constant
temperature and pressure, i.e., at thermodynamically well-defined
conditions. This holds true in particular for the small networks
discussed in cell biology where the few relevant molecules are
embedded in the aqueous intracellular solution. Such networks describing,
e.g., gene regulation \cite{mcad97, kepl01, paul04}, signal transduction (see, e.g. \cite{shib05}) or molecular motors \cite{parm99, fish99, lipo00, maes03a, bake04}, operate typically in
non-equilibrium generated by (time-dependent) external mechanical
or chemical stimuli. Feedback loops can generate cyles as a
characteristics of non-equilibrium steady states. Taking both the
fluctuation aspects and the well-defined thermodynamic
conditions seriously, the question arises
 whether or not it is possible to give a
thermodynamically consistent description of such networks on the
level of a single (stochastic) trajectory. 

Thermodynamics on the level of a fluctuating trajectory 
in a driven system may look 
far fetched at first sight. Such an approach, however, has been
extremly fruitful for somewhat different but conceptually related
systems consisting of a few degrees of freedom, like
colloidal particles or (bio)polymers \cite{bust05}. The dynamics of these systems
 in aqueous solution under the action of external forces generated
mechanically by the tip of an atomic force microscope or optically by
laser tweezers can be typically described by a Langevin equation.
For such a driven dynamics, a thermodynamically consistent description has
recently been worked out based on  two  essential ingredients which can
also be seen as the essential assumptions of such an approach.
(i) A first-law
like balance relates external work, internal energy and exchanged heat
dynamically along  each fluctuating trajectory \cite{seki98, shib00a}. 
(ii) The heat dissipated into the environment leads to a well-defined
entropy increase of the surrounding medium  which is assumed to keep
locally  its constant temperature
despite the fact that the few degrees of 
freedom of the system (colloidal particle or polymer)  embedded into this
bath are driven to non-equilibrium. For such systems, various exact relations 
have been derived theoretically. The most prominent 
examples are the Jarzynski relation \cite{jarz97, jarz97a, croo99, croo00} which allows 
to extract free energy
differences from experiment or simulations performed under non-equilibrium
conditions and the fluctuation theorem  \cite{evan94, gall95, kurc98, lebo99, maes03, maes03b} 
which quantifies the probability
to observe entropy annihilating trajectories in the steady state. These
relations and various ramifications \cite{croo00, hata01, seif05a} have been tested in
real and computer experiments both using
mechanical stretching of biomolecules \cite{humm01, liph02,  park04,  brau04, coll05, spec05} as well as colloidal particles
driven by laser traps \cite{wang02, carb04, trep04, blic06}. Since the fundamental 
constituents 
 (a stochastic trajectory,
a surrounding heat bath and some source of non-equilibrium) of these
mechanical systems are also present in chemical networks as described
above, 
a similar approach should be feasible for them as well. The purpose
of this paper is to develop this correspondence and explore its consequences
in some generality formally. 

Fluctuation theorems have indeed been discussed previously both for
the steady state of non-equilibrium networks by Gaspard and coworkers \cite{gasp04, andr04,andr06} and
for time-dependently driven chemical reactions in Ref.  \cite{seif04} without making the
connection to a first-law energy balance. Qian and co-workers have developed
such an energetic perspective mainly on the ensemble level \cite{qian01, qian05, qian05a, heue06} extending Hill's classical work \cite{hill}. On the trajectory level, unpublished work by Shibata \cite{shib00} and our previous studies on
simple models for enzyms \cite{schm06a} point in this direction.

The present systematic approach towards a stochastic thermodynamics \cite{mou86}
for chemical networks based on a master equation 
rests on two main ingredients.  First, we apply the first law to a single 
trajectory, i.e. we analyse a 
single reaction event
in terms of chemical work applied by the chemiostats, change in internal energy and 
amount of dissipated heat. This allows us to express the exchanged heat
in each reaction step in terms of the rate constants of the master equation. 
Second, we define entropy production also along a single trajectory
as a sum of two contributions. The heat dissipated in each step increases
the entropy of the surrounding medium. Moreover, there is an entropy of the
network itself. The sum of both changes obeys various exact relations
from which, e.g., the second law follows trivially for the mean entropy
production.
 
The paper is organized as follows. In Section \ref{model}, we recall the stochastic modelling of chemical reaction networks by a master equation approach. In Section \ref{sec3}, we define energy, work and dissipated heat on a stochastic trajectory and develop a consistent nonequilibrium first law along a single stochastic trajectory of reaction events. In Section \ref{sec4}, we define stochastic entropy along a trajectory of the system and total entropy production of system and heat bath. In Section \ref{sec5}, we derive a fluctuation theorem for the total entropy production from which a second-law like statement follows directly. We then relate this fluctuation theorem to the Jarzynski relation. The total entropy production can be split up into a housekeeping heat, which accounts for the irreversibility due to nonequilibrium steady states, and a quantity, which accounts for the irreversibility due to transitions between steady states \cite{oono98, hata01}. Both quantities obey a generalized fluctuation theorem. In Section \ref{sec6}, we apply our results to a paradigmatic reaction network.

\section{Model}
\label{model}

\subsection{Chemical master equation}

\begin{figure}
 \includegraphics[width = 0.5\linewidth]{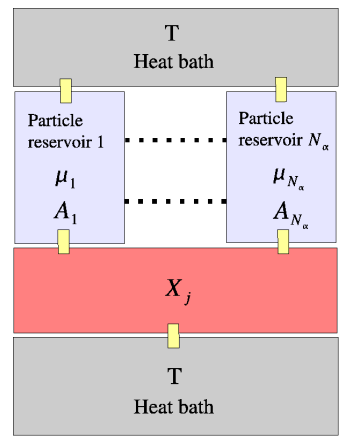} 
 \caption{Coupling of the system with species $X_j$, $j = (1,\dots,\Nj)$ to the $\Nal$ particle reservoirs for species $A_\al$ at chemical potential $\mu_\al$ and to a heat bath at constant temperature $T$.}
\label{coupling} 
\end{figure}

We consider the general reaction network \cite{vankampen, gardiner}
\beq
\sum_{\al=1}^{\Nal} \cla_\al^{\rho} A_\al + \sum_{j=1}^{\Nj} \clx_j^{\rho} X_j \rightleftharpoons  \sum_{\al=1}^{\Nal} \cra_\al^{\rho}A_\al +  \sum_{j=1}^{\Nj} \crx_j^{\rho}X_j  
\label{react_netw}
\eeq
with $1 \leq \rho \leq \Nrho$ labeling the single (reversible) reactions. We distinguish two types of reacting species. The $X_j$ molecules $(j = 1,\dots,\Nj)$ are those species whose numbers $\mathbf{n} = \left( n_1, \dots, n_{N_j}\right)$ can, in principle, be measured along a chain of reaction events. In practice, these numbers should be small. The $A_\al$ molecules $(\al = 1,\dots,\Nal)$ correspond to those species whose overall concentrations $c_\al$ are controlled externally by a chemiostat due to a (generally) time dependent protocol $c_\al(\tau)$. In principle, this implies that after a reaction event has taken place, the used $A_\al$ are ``refilled'' and the produced ones are ``extracted''. These chemiostats have chemical potential
\beq
\mu_\al = \mu^0_\al + T \ln \left( {c_\al}/{c_0} \right)
\label{def_mu_al_kon}
\eeq
where $\mu^0_\al$ is the chemical potential for standard conditions at concentration $c_0$ and $T$ is the temperature of the heat bath to which both type of particles are coupled, see Fig. \ref{coupling}. 
Throughout the paper we set Boltzmann's constant to unity which implies that entropy will be dimensionless. We assume that the reacting species have no internal degrees of freedom. However, internal degrees of freedom could easily be treated within our approach by labeling different internal states as different species and defining ``reactions'' (transitions) between them.

The stochiometric coefficients  $\cla_\al$, $\clx_j$, $\cra_\al$ and $\crx_j$ enter the stochiometric matrix $\mathbf{V}$ with entries
\beq
v_j^{\rho} \equiv \crx_j^{\rho}-\clx_j^{\rho} 
\eeq
and the stochiometric matrix of the reservoir species $\mathbf{U}$ with entries
\beq
u_\al^{\rho} \equiv \cra_\al^{\rho} - \cla_\al^{\rho} .
\eeq
For the externally controlled concentrations $c_\al$ of $A_\al$, we will use the vector notation $\ca = \left( c_1, \dots, c_{\Nal}\right)$ in the following. We assume a dilute solution of reacting species in a solvent (heat bath) and therefore the transition probabilities per unit time for the $\Nrho$ reactions (\ref{react_netw}) take the text book form \cite{vankampen, gardiner}
\beq
w_{+}^{\rho}(\mathbf{n}, \ca) = \Omega  k_{+}^{\rho}\prod_\al \calph^{\cla_\al^{\rho}}\cdot \prod_j \frac{n_j! }{(n_j-\clx_j^{\rho})!\cdot \Om^{\clx_j^{\rho}}}   
\label{w+}       
\eeq
\medskip
\beq
w_{-}^{\rho}(\mathbf{n}, \ca) =  \Omega  k_{-}^{\rho}\prod_\al \calph^{\cra_\al^{\rho}}\cdot \prod_j \frac{n_j! }{(n_j-\crx_j^{\rho})!\cdot \Om^{\crx_j^{\rho}}}   
\label{w-}  
\eeq
where $+$ denotes a forward reaction, $-$ denotes a backward reaction and $\Omega$ is the reaction volume.  
The bare rates $k^{\rho}_{+,-}$ are the transition probabilities per unit time per unit volume per unit concentration (in terms of $1 / \dbv_\al$ and $1 / \dbv_j$, respectively) of every educt reactant. Note that while $w_{+,-}^{\rho}(\mathbf{n}, \ca)$, in principle, can be measured experimentally, the bare rates $k_{+,-}^{\rho}$ depend on the normalizing volumes $\dbv_\al$ and $\dbv_j$ whose unique definition, as we will see in Sect. \ref{int_energy}, requires a microscopic Hamiltonian. 
 
The transition probabilities depend only on the current state and therefore define a Markov process with a unique master equation
\begin{eqnarray}
\partial_{\tau} p(\mathbf{n},\tau) &=& \sum_{\rho} \left[ w_{+}^{\rho}(\mathbf{n- v^{\rho}}, \ca)\cdot p(\mathbf{n-v^{\rho}},\tau) \right. \nn \\ 
&&~~~~~ + \left. w_{-}^{\rho}(\mathbf{n+v^{\rho}}, \ca)\cdot p(\mathbf{n+v^{\rho}},\tau) \right] \nn \\
%\end {equation*}
%\beq
&-&\sum_{\rho} \left[ w_{+}^{\rho}(\mathbf{n}, \ca)\cdot p(\mathbf{n}, \tau) \right. \nn \\
&& ~~~~~\left. +w_{-}^{\rho}(\mathbf{n}, \ca)\cdot p(\mathbf{n},\tau) \right] 
%\eeq
\label{master_eq}
\end{eqnarray}
governing the time evolution of the probability distribution $p(\mathbf{n},\tau)$ to have $n_j$ molecules $X_j$ at time $\tau$. Here, we have used the vector notation $\mathbf{v}^{\rho} = (v_1^{\rho}, \dots, v_{\Nj}^{\rho})$ for the entries of the stochiometric matrix.

\subsection{Stationary states}

Solving the time-dependent master equation (\ref{master_eq}) is out of scope for most systems. For constant $c_\al$, the long-time limit of any initial distribution is a stationary state under quite general conditions \cite{schn76}. The master equation for the relaxation process (at constant $c_\al$) can be solved analytically for linear reaction networks \cite{heue06}. We first explore whether any given network allows for an equilibrium solution $p^{\eq}(\n,\ca)$ which obeys the detailed balance condition
\beq
p^{\eq}(\mathbf{n}, \ca) w_{+}^{\rho}(\mathbf{n}, \ca) = p^{\eq}(\mathbf{n + v^{\rho}}, \ca) w_{-}^{\rho}(\mathbf{n + v^{\rho}}, \ca).
\label{det_bal}
\eeq
Using (\ref{w+}) and (\ref{w-}), we get 
\beq
\frac {p^{\eq}(\mathbf{n}, \ca)} {p^{\eq}(\mathbf{n + v^{\rho}}, \ca)} =  \frac {k_-^{\rho}} {k_+^{\rho}} \prod_\al \calph^{u_\al^{\rho}} \prod_j \frac {(n_j + v_j^{\rho})! \Omega^{v_j^{\rho}} }{{n_j!} \dbv_j^{v_j^{\rho}}  }.
\label{eq_distr_cond}
\eeq
Inserting a multivariate Poissonian 
\beq
p^{\eq}(\mathbf{n}, \ca) = \prod_j \frac {1} {n_j!} \left ( {n_j^s} \right)^{n_j} e^{-n_j^s}
\label{eq_distr}
\eeq
with mean $n_j^s$ into equation (\ref{eq_distr_cond}) yields the $\Nrho$ equations \cite{gardiner}
\beq
\frac {k_+^{\rho}}{k_-^{\rho}} =   \prod_\al \calph^{u_\al^{\rho}} \prod_j \left (\frac {n_j^s \dbv_j}{\Omega} \right)^{v_j^{\rho}}.
\label{ns_equation}
\eeq
These $\Nrho$ equations (\ref{ns_equation}) are equivalent to the system of linear equations
\beq
\sum_j v_j^{\rho} \epsilon_j = \ln \frac {k_-^{\rho}} {k_+^{\rho}} + \sum_\al u_\al^{\rho} \ln \calph
\label{epsilonj}
\eeq
with 
\beq
\epsilon_j \equiv - \ln \left( n_j^s \dbv_j / \Omega \right). 
\eeq
We now have to distinguish three cases concerning the solvability of the system (\ref{epsilonj}) which lead to three different physical situations.

\subsubsection*{Case I : Equations (\ref{epsilonj}) have a unique solution. }
Generically, this case requires $\Nrho = \Nj$. The unique solution of eqs. (\ref{epsilonj}) corresponds to the unique equilibrium distribution, if all states are accessible from every initial state through a sequence of reaction events.
The equilibrium distribution (\ref{eq_distr}) can then be written as
\beq
p^{\eq}(\mathbf{n}, \ca) = \mathcal N \prod_j \frac {1} {n_j!} \Om^{n_j} e^{-\epsilon_j n_j} 
\label{eq_distr1}
\eeq
with normalization factor $ \mathcal N = \prod_j e^{-n_j^s}$. Taken at face value, the form (\ref{eq_distr1}) resembles a grandcanonical distribution with one particle energies $\epsilon_j$ and chemical potentials $\mu_j = 0$. As we will show below, consistency requires splitting $\epsilon_j$ in a one-particle energy $E_j$ and a non-zero chemical potential $\mu_j$.

However, even if eqs. (\ref{epsilonj}) have a unique solution, not all states must necessarily be accessible from every initial state through a sequence of reactions due to the discrete reaction kinetics. The equilibrium distribution then is the projection of the Poissonian (\ref{eq_distr1}) to the subspace of accessible states. 

\subsubsection* {Case II : Equations (\ref{epsilonj}) have more than one solution. }
The rank of the stochiometric matrix $\mathbf{V}$ then is smaller than the number $\Nj$ of species $X_j$. This is usually the case if there are less reactions $\rho$ than species $X_j$, i.e. $\Nrho < \Nj$. We then have constraints for the accessible state space depending on the initial distribution. The state space $\{\mathbf{n}\}$ is then separated into an infinite number of subspaces. If all possible initial states lie in the same subspace, the equilibrium distribution is the projection of the Poissonian (\ref{eq_distr1}) to this subspace with the $\epsilon_j$ given by any solution of (\ref{epsilonj}). As is shown in App. \ref{App1}, it does not matter which solution of (\ref{epsilonj}) is chosen.

\subsubsection*{Case III : Equations (\ref{epsilonj}) have no solution. }
The chosen $c_\al$ then create a genuine nonequilibrium stationary state (NESS) violating the detailed balance condition. This will usually happen, if there are more reactions $\rho$ than species $X_j$, i.e., $\Nrho > \Nj$. We then have essentially different paths leading to the same final state, i.e., there are genuine cycles in the network.

Even if the network has no equilibrium solution for the given $c_\al$, there must be certain values $c_\al^{\eq}$ for which eqs. (\ref{epsilonj}) have a solution. For we can always assume to isolate the system after preparing an initial state with given $c_\al$. The network will then relax into an equilibrium state with concentrations $c_\al^{\eq}$.

\label{cases}
\section{First law}
\label{sec3}

We want to state the first law of thermodynamics along a stochastic trajectory $\n(\tau)$ for an arbitrary network (\ref{react_netw}). We therefore need the concept of internal energy, chemical work and dissipated heat.

\subsection{Internal energy}
We start with internal energy and consider an equilibrium situation with concentrations $c_\al^{\eq}$ which are even defined for case III, as discussed above. Then, a solution of (\ref{epsilonj}) exists for this set of reservoir concentrations $\ceq_\al$. We also assume a microscopic Hamiltonian for the dilute solution of $X_j$ and $A_\al$ molecules in the solvent from which an (average) energy $E_j$ and $E_\al$ for the reacting species $X_j$ and $A_\al$, respectively, can be derived as shown in App. \ref{App2}. 
The chemical potential (\ref{def_mu_al_kon}) then can be written as
\beq
\mu_\al = E_\al + T \ln \calph
\label{def_mu_al}
\eeq 
where the volumes $\dbv_\al$ are chosen such that $E_\al$ becomes the energy of one molecule $A_\al$ as defined by a microscopic Hamiltonian, see App. \ref{App2} for details. If, for example, the $A_\al$ could be treated as an ideal gas without internal degrees of freedom, this normalization volume would become $\dbv_\al \equiv \lambda_\al^3 \exp({-3/2})$ where $\lam_\al$ is the thermal de Broglie wavelength of an $A_\al$ molecule and the factor $\exp(-3/2)$ guarantees that the thermal kinetic energy is included in the energy $E_\al$ of an $A_\al$ molecule \cite{mcquarrie}. 

Even though the rewriting from eq. (\ref{def_mu_al_kon}) to eq. (\ref{def_mu_al}) looks trivial, it is a crucial step since it provides the relation between the energy $E_\al$ which will enter the first law and the chemical potential $\mu_\al$. The very fact that there is a concentration $c_\al = 1 / \dbv_\al$ where both become numerically identical does, of course, not imply that these two quantities are similar in general. In fact, the volume factor $\dbv_\al$ as derived in  App. \ref{App2} depends on the microscopic details.

The key point for the subsequent analysis is that the energies $E_j$ thus identified in an equilibrium situation from statistical mechanics are independent from $\ca^{\eq}$. Since in a dilute solution, there is no interaction between the reacting species apart from the chemical reactions, the energy $E_j$ of one molecule $X_j$ will neither depend on the number of molecules $n_i$ of the other species nor on the concentrations $c_\al$. Hence, the $E_j$ are well defined even for networks in a NESS and for networks driven by time-dependent $c_\al(\tau)$. Likewise, the definition of the chemical potential of reservoir species (\ref{def_mu_al}) still holds true in such nonequilibrium situations, since the particle reservoirs are assumed to be large. 

A system energy or internal energy 
\beq
E(\mathbf{n(\tau)}) = \sum_j E_j n_j(\tau)
\eeq
along a stochastic trajectory can then be defined both in equilibrium and under the specified non-equilibrium conditions. 

\label{int_energy}

\subsection{Work and heat}

For a formulation of the first law along a stochastic trajectory, we next have to calculate the heat flowing from the system into the heat bath. Whenever a reaction $\rho$ takes place in forward direction ($r \equiv +1$) or backward direction ($r \equiv -1$), i.e., whenever $\mathbf{n}$ changes, the energy of the system changes by
\beq
\Delta E^{r, \rho} \equiv \Delta E^{\pm, \rho} \equiv \sum_{j} E_{j} \Delta n_{j}^{r, \rho} = \pm \sum_{j} E_{j} v^{\rho}_{j}.
\label{dE_single}
\eeq
where $\Delta n_j^{r, \rho}  \equiv r v_\al^{\rho}$ is the change in the number of $X_j$ molecules due to the reaction $r , \rho$.
The chemical work done by the particle reservoirs in such a reaction step is 
\beq
\Wc^{r, \rho} \equiv \Wc^{\pm, \rho} = -\sum_\al \mu_\al \Delta n_\al^{r, \rho} = \mp \sum_\al \mu_\al u_\al^{\rho}. 
\label{W_single}
\eeq
where $\Delta n_\al^{r, \rho}  \equiv r u_\al^{\rho}$ is the number of $A_\al$ molecules transformed in the reaction $\pm \rho$. Due to energy conservation, the heat flowing from the system into the heat bath follows as 
\beq
Q^{r, \rho} \equiv \Wc^{r, \rho} - \Delta E^{r, \rho}  = - \sum_\al \mu_\al \Delta n_\al^{r, \rho} - r \sum_{j} E_{j} v^{\rho}_{j} .
\label{Q_single}
\eeq
Thus, we have identified the first law of thermodynamics for a single reaction event as
\beq
\Wc^{r, \rho} = \Delta E^{r, \rho} +  Q^{r, \rho}.
\label{first_law}
\eeq

For a finite time intervall $[0, t]$ with given concentration protocols $c_\al(\tau)$, we sum (\ref{dE_single}), (\ref{W_single}) and (\ref{Q_single}) over all occuring reactions $\rho_l$ at times $\tau_l$, where we will denote forward reactions by $r_l = +1$ and backward reactions by $r_l = -1$. We then get the chemical work
\begin{eqnarray}
\Wc &\equiv& \sum_l W^{r_l, \rho_l} =  - \sum_l  \sum_\al \mu_\al(c_\al(\tau_l)) \Delta n_\al^{r_l , \rho_l} \nn \\
&=&  -\int_0^t d\tau \left ( \sum_\al \mu_\al(c_\al(\tau)) \dot n_\al \right),
\label{W}
\end{eqnarray}
the change in internal energy
\beq
\Delta E = \int_0^t d\tau \left ( \sum_j E_j \dot n_j \right)
\eeq
and the dissipated heat
\begin{eqnarray}
Q &=& -\Delta E - \sum_l  \sum_\al \mu_\al(c_\al(\tau_l)) \Delta n_\al^{r_l , \rho_l} \nn \\&=& - \int_0^t d\tau \left (\sum_j E_j \dot n_j + \sum_\al \mu_\al(c_\al(\tau)) \dot n_\al \right)
\end {eqnarray}
with 
\beq
\dot n_\al \equiv \sum_l \delta(\tau - \tau_l) r_l u_\al^{\rho_l}
\eeq
and 
\beq
\dot n_j \equiv \sum_l \delta(\tau - \tau_l) r_l v_j^{\rho_l}.
\eeq
Since only energies $E_j$ and chemical potentials $\mu_\al$ enter the crucial quantities $\Delta E$, $Q$ and $W$, this formulation of the first law is also valid in nonequilibrium.

The present identification of work, internal energy and heat depends crucially on the fact that we control the concentrations $c_\al$ externally. This means in particular that the chemical work has to be spent for ``refilling'' the chemiostats after each reaction step. A somewhat different identification applies if one considers relaxation (without further external interference) from an initially prepared nonequilibrium state with concentrations $c_\al(\tau = 0)$ to the corresponding equilibrium with concentrations $\ceq_\al$ for $\tau \to \infty$. In this case, the system should comprise the $X_j$ and the $A_\alpha$ molecules. Then we cannot distinguish between the two types of species $X_j$ and $A_\alpha$ anymore and we have to label all reacting species with $\tilde X_{\tilde j}$, $(\tilde j = 1, \dots, \Nal + \Nj)$ leaving no chemiostat species $\tilde A_\al$. The change in the internal energy of the system for a single reaction step then is 
\beq
{\Delta \tilde E} \equiv \sum_{\tilde j} \tilde v_{\tilde j}^{\rho} \tilde E_{\tilde j} \equiv \sum_j v_j^{\rho} E_j + \sum_\al u_\al^{\rho} E_\al
\eeq
and since the first law then involves no external chemical work anymore, $Q \equiv - {\Delta \tilde E}$ is dissipated as heat. Our results (including those of the following sections) hold for this special case, too, if they are applied to the system of $\tilde X_{\tilde j}$ without chemiostat species.

\subsection{Relation between statistical and kinetic approach}
In equilibrium, the chemical potentials $\mu_j$ must obey
\beq
\sum_j v_j^{\rho} \mu_j \equiv  -\sum_\al u_\al^{\rho} \mu_\al =    -\sum_\al u_\al^{\rho}  \left[ E_\al +  T \ln \left( \ceq_\al \dbv_\al \right) \right]
\label{def_mu_j}
\eeq
since there is no chemical potential difference between connected states in equilibrium. We now compare the equilibrium distribution (\ref{eq_distr1}) to the grandcanonical equilibrium distribution of a dilute solution, which can always be written as (see App. \ref{App2} for details)
\beq
p^{\eq}(\mathbf{n}, \caeq) = \mathcal N \prod_j \frac {\Om^{n_j}} {n_j!} e^{-\beta (E_j n_j - \mu_j n_j)}
\label{grandcan_distr}
\eeq
with the energy $E_j$ of one $X_j$ molecule and appropriate normalization volumes $\dbv_j$.  
Using the relations for the chemical potentials $\mu_j$ (\ref{def_mu_j}) and the system of linear equations for the $\epsilon_j$ (\ref{epsilonj}), we see that thermodynamic consistency constrains the ratio of bare rates ${k_-^{\rho}}/{k_+^{\rho}}$ by
\beq
\beta \sum_j v_j^{\rho} E_j = \ln \frac {k_-^{\rho}} {k_+^{\rho}} - \beta \sum_\al u_\al^{\rho}  E_\al
\label{def_E_j}
\eeq
where $\beta \equiv 1 /  T$ is the inverse temperature. 

Two remarks are appropriate. First, in general the energies $E_j$ cannot be extracted from the kinetics, i.e. the chemical master equation, without knowledge of the detailed Hamiltonian since the bare rates $k_{+,-}^{\rho}$ depend on the normalizing volumes $\dbv_\al$ and $\dbv_j$ which are defined only through a microscopic Hamiltonian.
Second, if one assumes the $k_{+,-}^{\rho}$ and the $\dbv_\al$, $\dbv_j$ to be given without specifying an underlying microscopic Hamiltonian, one has to make sure that the energies $E_\al$ are chosen such that eqs. (\ref{def_E_j}) have a solution for the $E_j$. From another viewpoint, given energies $E_\al$ and $E_j$ constrain the possible bare rates for a given thermodynamically consistent reaction network. An illustration of this fact will be given in Sect. \ref{sec6} below.
 
\label{rel}

\section{Entropy}
\label{sec4}

We distiguish an entropy change of the system proper from a change in the entropy of the medium (heat bath). The total entropy production is the sum of both entropy changes. 

\subsection{Medium entropy change}

The dissipated heat $Q$ from the first law corresponds to a change in entropy of the heat bath
\beq
\Delta \sm \equiv  \frac Q T  =  - \frac 1 T \int_0^t d\tau \left ( \sum_j E_j \dot n_j + \sum_\al \mu_\al \dot n_\al \right)
\label{Q}
\eeq
along a single trajectory. 
The change in medium entropy due to one occuring reaction $\rho$ in direction $r$ can be expressed through rate constants using the definitions of energies (\ref{def_E_j}) and chemiostat chemical potentials (\ref{def_mu_al}) as
\begin {eqnarray}
\Delta \sm^{r, \rho} &\equiv& \frac 1 T Q^{r, \rho} = - r \beta \left( \sum_{j} E_{j} v^{\rho}_{j} + \sum_\al \mu_\al u_\al^{\rho} \right) \nn \\
&=& r \ln \frac {k_+^{\rho}} {k_-^{\rho}} - r \sum_\al u_\al^{\rho} \ln \calph \nn \\
&=&  \ln \frac {w_{r}^{\rho}(\mathbf{n}, \ca)}{w_{-r}^{\rho}(\mathbf{n} + r \mathbf{v^{\rho}}, \ca)} - \sum_j \ln \frac {\Om^{r v_j^{\rho}} n_j!} {  (n_j + r v_j^{\rho})!} .
\label{Q_rates}
\end {eqnarray}
In the third line, we have used the transition probabilities (\ref{w+}) and (\ref{w-}).
Summing (\ref{Q_rates}) over all occuring reactions  $\rho_l$ in direction $r_l$ at times $\tau_l$ yields the medium entropy change
\beq
\Delta \sm = \sum_l r_l \ln \frac {k_+^{\rho_l}} {k_-^{\rho_l}} - \sum_l r_l \sum_\al u_\al^{\rho} \ln \calph
\label{med_entr}
\eeq
along a single trajectory.

\subsection{System entropy}

We will now define an entropy of the system. The total entropy production of system and heat bath then is the relevant quantity for the second law of thermodynamics. Again, we will be guided by the equilibrium case first.
Equilibrium state functions can be obtained from (\ref{grandcan_distr}) using the partition function
\begin{eqnarray}
Z &=& \sum_{\mathbf{n}} \prod_j \frac {\Om^{n_j}} {n_j!} e^{-\beta (E_j n_j - \mu_j n_j)} \nn \\ 
&=&  \prod_j \exp \left (\Om e^{-\beta  (E_j - \mu_j)} \right)
\end{eqnarray}
and the grandcanonical potential
\beq
J = - T \ln Z = - T \sum_j \Om e^{-\beta  (E_j - \mu_j)} .
\eeq
Of particular importance is the equilibrium entropy of the system 
\begin{eqnarray}
S^{\eq} &=& -\partial_T J \nn \\
&=&  -  \sum_{\mathbf{n}} p^{\eq}(\mathbf{n})\left( \ln p^{\eq}(\mathbf{n}) - \sum_j \ln \frac{\Om^{n_j}}{ n_j!} \right) .
\label{entropy_therm}
\end{eqnarray}
This expression differs by a term
\beq
g_{\mathbf{n}} = \prod_j \frac {\Om^{n_j}} {n_j!}
\label{deg_fac}
\eeq
involving the degeneracy of the state $\n$ from the usual Shannon entropy 
\beq
S_{\rm{Sh}} \equiv - \sum_{\mathbf{n}} p(\mathbf{n}) \ln p(\mathbf{n}) = \mean{- \ln p(\mathbf{n})}
\eeq
of a Markov process. Relaxing the equilibrium constraint, we define the ensemble entropy in non-equilibrium as
\begin{eqnarray}
S(\tau) &\equiv& -\sum_{\mathbf{n}} p(\mathbf{n}, \tau)\left( \ln p(\mathbf{n}, \tau) - \ln g_{\n} \right)  \nn \\
&=&  \mean{- \ln p(\mathbf{n}, \tau) + \ln g_{\mathbf{n}} }.
\label{eq_entropy}
\end{eqnarray}

We will now generalize this ensemble expression to a stochastic entropy valid along a single trajectory for both equilibrium and nonequilibrium situations following \cite{seif05a}.  
We define the stochastic system entropy
\beq
s(\tau) \equiv -\ln p(\mathbf{n}(\tau), \tau) + s^0(\mathbf{n(\tau)}),
\label{system_entropy}
\eeq
where $p(\mathbf{n}(\tau),\tau)$ is the solution of the master equation (\ref{master_eq}) taken along the stochastic trajectory $n(\tau)$ and
\beq
s^0(\mathbf{n(\tau)}) \equiv \ln g_{\mathbf{n(\tau)}} \equiv \sum_j\ln \frac  {\Om^{n_j(\tau)}} {n_j(\tau)!}
\eeq
is the internal entropy of the state $\mathbf{n}$ due to the degeneracy. The ensemble entropy (\ref{eq_entropy}) then is the average of the stochastic entropy (\ref{system_entropy}).  

\subsection{Total entropy production}

The total entropy production along a stochastic trajectory 
\begin{eqnarray}
\Delta \stot &=& \Delta s + \Delta \sm \nn \\
&=& \ln \frac {p(\n(0),0)}{p(\n(t), t)} + \sum_j \ln \frac {\Om^{n_j(t)} n_j(0)!}{\Om^{n_j(0)} n_j(t)!} \nn \\ &+& \sum_l r_l \ln \frac {k_+^{\rho_l}} {k_-^{\rho_l}} - \sum_l r_l \sum_\al u_\al^{\rho} \ln \calph
\label{dstot}
\end{eqnarray}
then is the sum of the changes of system entropy (\ref{system_entropy}) and medium entropy (\ref{med_entr}),
where $\Delta s \equiv s(t) - s(0)$. It will be shown in Sect. \ref{Sect_FT} below that, in fact, $\Delta \stot$ is independent of the normalizing volumes $\dbv_\al$ and $\dbv_j$. Hence, this quantity does not require explicitly a microscopic Hamiltonian but is rather determined by the full rates $w_{+,-}^{\rho}(\mathbf{n}, \ca)$ entering the master equation. As an aside, we note that after averaging the stochastic entropy production $\Delta \stot$, the ensemble entropy production from Ref. \cite{gasp04} is recovered.

\subsection{Stochastic Gibbs relation}

We can now relate the identified first law (\ref{first_law}) to the usual equilibrium Gibbs relation. Inserting the definition of the medium entropy (\ref{Q}) into (\ref{first_law}), we find
\beq
\Delta E = -T \Delta \sm - \sum_\al \mu_\al \Delta n_\al
\eeq
In equilibrium and for quasistatic transitions, we have $\Delta \stot = 0$ and therefore $\Delta s = - \Delta \sm$ which implies
\beq
\Delta E = T \Delta s + \sum_j \mu_j \Delta n_j.
\eeq
Specialized to equilibrium, our approach thus yields along a single stochastic trajectory the usual thermodynamic relations valid for ensemble averages in equilibrium.

\section{Fluctuation theorems}
\label{sec5}

Fluctuation theorems provide exact results for driven systems beyond linear response. So far, they have been studied extensively for Langevin dynamics and for master equations in general. Previous applications to chemical reaction networks \cite{gasp04, andr04, seif04} have neither made the connection to an energy balance nor considered the entropy on the stochastic level with occasional exceptions \cite{shib00, seif05, seif05a, schm06a}. The purpose of this section is to develop this concept in full generality.

\subsection{Integral fluctuation theorem}

The weight $p[\n(\tau), \ca(\tau)]$ for a particular trajectory $\n(\tau)$ with given initial state $\n_0$ under the concentration protocol $\ca(\tau)$ is given by  
\begin {eqnarray}
p[\n(\tau), \ca(\tau)] &=& \prod_{l=1}^{N_l+1} \exp \left[{-\sum_{r, \rho} \int_{\tau_{l-1}}^{\tau_{l}} d\tau   w_r^{\rho}(\mathbf{n}_{l}^-, \ca(\tau))  } \right] \times \nn \\
&&\times \prod_{l=1}^{N_l}  {w_{r_l}^{\rho_l}(\mathbf{n}^-_{l}, \ca(\tau_l) )} ,
\label{traj_weight_M}
\end{eqnarray}
with $\tau_l$ ($l=1,\dots,N_l$) being the jump times, where a reaction $\rho_l$ in direction $r_l = +,-$ from state $\mathbf{n}^-_{l}$ to state 
\beq
\mathbf{n}^+_l \equiv \mathbf{n}^-_l \pm \mathbf{v}^{\rho_l}
\eeq
occurs. For ease of notation, we have introduced the initial time $\tau_0 \equiv 0$, the final state $\n^-_{N_l+1} \equiv \n^+_{N_l}$ and the final time $\tau_{N_{l+1}} \equiv t$. We then define
 \begin{equation}
R \equiv \ln \frac {p_0(\n_0) \cdot p[\n(\tau), \ca(\tau)]} {p(\n_t, t) \cdot p[\tilde{\n}(\tau), \tilde {\ca}(\tau)]}
\label{R_allg}
\end{equation}
as the logarithm of the ratio of trajectory weights for a given trajectory $\n(\tau)$ with initial state $\n_0 \equiv \n(0)$ and final state $\n_t \equiv \n(t)$ and the corresponding trajectory $\tilde \n (\tau) \equiv \n(t-\tau)$ under the time reversed protocol $\tilde{\ca}(\tau) \equiv \ca (t-\tau)$.  Here, $p_0(\n_0)$ is the initial distribution and $p(\n_t, t)$ is the solution of the master equation at time $t$. 
The integral fluctuation theorem 
 \beq
\left\langle e^{-R}\right\rangle = 1
\label{FT}
\eeq
can then be proven by the following lines of identities \cite{lebo99, croo99, maes03, seif05a} using the normalization condition for the trajectory weight (\ref{traj_weight_M}) and the final distribution $p(\n_t, t)$
\begin{eqnarray}
1 &=& \sum_{\tilde \n (\tau)} p(\n_t, t) p[\tilde \n (\tau), \tilde {\ca}(\tau)] \nn  \\
 &=& \sum_{\tilde \n (\tau)} e^{-R[\n(\tau)]} p_0(\n_0) p[\n (\tau), \ca(\tau)]  \nn \\
 &=& \sum_{\n (\tau)} e^{-R[\n(\tau)]}  p_0(\n_0) p[\n (\tau), \ca(\tau)] \nn \\
 &=& \mean { e^{-R[\n(\tau)]}}.
\end{eqnarray} 

Our thermodynamic approach allows to show that the abstract quantity $R$ is exactly the total entropy production $\Delta \stot$.  
Inserting the trajectory weights (\ref{traj_weight_M}) into (\ref{R_allg}) yields
\begin{eqnarray}
R &\equiv& \ln \frac {p_0(\n_0) p[\n(\tau),\ca(\tau)]}{p(\n_t, t) p[\tilde{\n}(\tau), \tilde {\ca}(\tau)]} \nn \\ 
&=& \ln\frac {p_0(\n_0)} {p(\n_t, t)} + \sum_{l} \ln \frac {w_{r_l}^{\rho_l}(\mathbf{n}^-_{l}, \ca_l)}{w_{-r_l}^{\rho_l}(\mathbf{n}^+_{l}, \ca_l)},
\label{R_rates}
\end{eqnarray}
where $\ca_l \equiv \ca(\tau_l)$. We then use (\ref{Q_rates}) to get
\begin{eqnarray}
R &=& \ln\frac {p_0(\n_0)} {p(\n_t, t)} + \sum_l \ln \left ( \frac {\n^-_l !} {\n^+_l!} \Oom^{\n^+_l - \n^-_l} \right) \nn \\
&&- \sum_l  \ln \left ( \frac {\n^-_l !}{\n^+_l!} \Oom^{\n^+_l - \n^-_l} \right) + \sum_{l} \ln \frac {w_{r_l}^{\rho_l}(\mathbf{n}_l^{-}, \ca_l)}{w_{-r_l}^{\rho_l}(\mathbf{n}_l^{+}, \ca_l)} \nn \\
&=& \Delta s + \Delta \sm = \Delta \stot.
\label{Reqdstot}
\end{eqnarray}
Here, we have used the shorthand notations
\beq
\n ! \equiv \prod_j n_j!
\eeq
and
\beq
\Oom^{\n} \equiv \prod_j \Om^{n_j}.
\eeq
The key point in this analysis is the cancellation of the degeneracy terms from the change in system entropy $\Delta s$ with those from the entropy change of the medium $\Delta \sm$ in eq. (\ref{Reqdstot}). Thus, the total entropy production $\Delta \stot = R$ depends only on inital and final distribution $p_0(\n_0)$, $p(\n_t, t)$ and transition probabilities $w_{+,-}^{\rho}(\mathbf{n}, \ca)$, see eq. (\ref{R_rates}), which can, in principle, be extracted from experiments. Explicit knowledge of the normalizing volumes $\dbv_\al$ and $\dbv_j$ defined only via the microscopic Hamiltonian is not required.

The fluctuation theorem 
\beq
\mean{ e^{-\Delta \stot}} = 1
\label{FT_stot}
\eeq
then constrains the probability distribution of the total entropy production $\Delta \stot$ from which the second law 
\beq
\langle \Delta \stot \rangle~\geq~0
\label{meanRgeq0}
\eeq
follows directly via the Jensen inequality $\mean{e^x} \geq e^{\mean{x}}$. The relation (\ref{FT_stot}) holds for any initial state, any time-dependent protocol $c_\al(\tau)$ and any length $t$ of trajectories. Similar relations have been derived within a Hamiltonian dynamics in Ref. \cite{jarz99}.

\label{Sect_FT}

\subsection {Nonequilibrium steady states and detailed fluctuation theorem}

Likewise, for nonequilibrium steady states with constant $c_\al$, the detailed fluctuation theorem
\beq
p(-\Delta \stot) = e^{-\Delta \stot} p(\Delta \stot)
\eeq
can be shown exactly in the same fashion \cite{seif05a}. Here, $p(\Delta \stot)$ is the probability density function for the total entropy production. For the validity of this relation in the steady state for finite length $t$ of trajectories, the inclusion of the stochastic entropy of the system $\Delta s$ is crucial. In the long run, $\Delta \sm$ scales as $t$ whereas $\Delta s$ remains bounded.

\subsection{Generalized Jarzynski relation}

The Jarzynski relation \cite{jarz97, jarz97a} 
\beq
\mean{e^{-\beta W}} = e^{-\beta \Delta F}
\label{jarz_rel}
\eeq
expresses the free energy difference $\Delta F$ between two equilibrium states with a non-equilibrium average of the work $W[x(\tau)]$ spent in this transition, which is a functional of the system trajectory $x(\tau)$. In the form (\ref{jarz_rel}), it is valid only if the work is defined as
\beq
W = \sum_k \int_0^t \frac {d E}{d \lambda_k} \dot \lambda_k
\eeq
where the time-dependent protocols $\lambda_k(\tau)$ (equivalent to the $c_\al(\tau)$ in our context) determine the transition rates via the time-dependent potential $E(\lambda_k(\tau), x)$. This definition of work is usually appropriate in a canonical ensemble. We now derive a generalized version of the Jarzynski relation for our grandcanonical system. We consider transitions between equilibrium states and rewrite the quantity $\Delta \stot$ using (\ref{first_law}) and (\ref{Q}) as
\begin{eqnarray}
\Delta \stot &=& \Delta s + \beta Q = \Delta s + \beta \Wc - \beta \Delta E  \nn \\
 &=& \beta \left(\Wc + \Delta \left ( \sum_j \mu_j n_j \right) - \Delta J \right)
\label{R_Aufsp_Jarz}
\end{eqnarray}
where we have used the definition of the grandcanonical potential
\beq
\Delta J = \Delta E - T \Delta s + \sum_j \mu_j \Delta n_j .
\eeq
We now show that this change in the grandcanonical potential does not depend on initial ($\n_0$) and final ($\n_t$) state of the trajectory but only on the initial and final values of $\ca$. Using the grandcanonical equilibrium distribution (\ref{grandcan_distr}) and the definition of system entropy (\ref{system_entropy}) we get
\begin {eqnarray}
J &=& E - T s - \sum_j \mu_j n_j \nn \\
 &=& \sum_j (E_j - \mu_j) n_j +   T \ln p^{\eq}(\n, \ca) - \ln g_{\n} + \ln \mathcal N(\ca)\nn \\
&=& \sum_j (E_j - \mu_j) n_j - \sum_j (E_j n_j - \mu_j n_j) + \ln \mathcal N(\ca) \nn \\
&=&  \ln \mathcal N(\ca)
\end {eqnarray}
and therefore $\Delta J = \ln {\mathcal N(\ca_t)} - \ln{\mathcal N(\ca_0)}$.
Using (\ref{R_Aufsp_Jarz}), the fluctuation theorem turns into the generalized Jarzynski relation
\beq
\mean{e^{-\beta \left[\Wc - \Delta \left( \sum_j \mu_j n_j \right)\right]}} = e^{-\beta \Delta J}.
\label{jarz_rel_gen}
\eeq
Two remarks on the generality of this relation are appropriate.

(i) At constant $c^{\eq}_\al$ belonging to an equilibrium state, the exponent on the lhs of eq. (\ref{jarz_rel_gen}) can be transformed to 
\begin{eqnarray}
&&\Wc - \Delta \left ( \sum_j \mu_j n_j \right ) \nn \\
&&~=  -\int_0^t d\tau \left ( \sum_\al \mu_\al(c^{\eq}_\al) \dot n_\al \right) - \Delta \left ( \sum_j \mu_j n_j \right ) \nn \\
&&~= \int_0^t d\tau \left ( \sum_j \mu_j(c^{\eq}_\al) \dot n_j \right) - \Delta \left ( \sum_j \mu_j n_j \right ) \nn \\
&&~= -\int_0^t d\tau \left ( \sum_j \dot \mu_j (c^{\eq}_\al ) n_j \right )= 0
\label{exp_jarz_gen}
\end {eqnarray}
using the definitions of chemical work $\Wc$ (\ref{W}) and chemical potential $\mu_j$ (\ref{def_mu_j}). Thus, the result (\ref{jarz_rel_gen}) stays valid even if we do not await relaxation to the final equilibrium state. Such final relaxation implicitly assumed in the derivation of (\ref{jarz_rel_gen}) does not contribute to the exponent on the rhs.

(ii) If we could clamp the initial and final number of molecules $\n_0$ and $\n_t$, we would get the original relation
\beq
\mean{e^{-\beta W}} = e^{-\beta \Delta F}.
\eeq

\subsection{Generalized Clausius inequality}

As mentioned above, there are two types of nonequilibrium in chemical reaction networks : (i) a genuine NESS and (ii) any type of time-dependent $c_\al(\tau)$. It is useful to split 
\beq
\Delta \stot =  \ln\frac {p_0(\n_0)} {p(\n_t, t)} + \sum_{l} \ln \frac {w_{r_l}^{\rho_l}(\mathbf{n}^-_{l}, \ca_l)}{w_{-r_l}^{\rho_l}(\mathbf{n}^+_{l}, \ca_l)} \equiv Y + \beta \Qhk 
\eeq 
into a ($c_\al$-dependent) part 
\beq
Y \equiv \ln\frac {p_0(\n_0)} {p(\n_t, t)} + \sum_{l} \frac {p^s(\n_l^+, \ca_l)}{p^s(\n_l^-, \ca_l)}
\eeq
which accounts for non-adiabatic transitions and the so called housekeeping heat \cite{hata01, shib00, oono98} 
\beq
\beta \Qhk \equiv \sum_{l} \ln \frac {w_{r_l}^{\rho_l}(\mathbf{n}^-_{l}, \ca_l ) p^s(\n_l^-, \ca_l)}{w_{-r_l}^{\rho_l}(\mathbf{n}^+_{l}, \ca_l) p^s(\n_l^+, \ca_l)},
\eeq
which is the permanently dissipated heat in a nonequilibrium steady state for constant $c_\al$. 
Obviously, $\Qhk = 0$ if the detailed balance condition (\ref{det_bal}) holds for the stationary state $p^s(\n, \ca(\tau))$ for any time $\tau$. For systems in a steady state, $Y$ vanishes.
For transitions between steady states, i.e., starting and ending in a stationary state, the quantity $Y$ can be transformed to
\beq
Y = \int_0^t d\tau \sum_\al \pd {\Phi} {c_\al} \dot c_\al,
\eeq
with $\Phi \equiv - \ln p^s$.
It has been shown \cite{hata01, shib00}, that the quantity $Y$ obeys the integral fluctuation theorem (Hatano-Sasa-relation)
\beq
\mean{e^{-Y}} = 1.
\label{HS}
\eeq
From (\ref{HS}) we get a generalized Clausius inequality even for transitions between nonequilibrium steady states
\beq
\mean Y \geq 0
\eeq
or
\beq
\beta \mean{\Qex} \geq - \mean{\Delta s}
\label{Cl_eq}
\eeq
with the excess heat
\beq
\beta \Qex \equiv  \Delta \sm - \beta \Qhk = \sum_{l=1}^{k-1} \frac {p^s(\n_l^-, \ca_l) g_{\n_l^-}}{p^s(\n_l^+, \ca_l) g_{\n_l^+}} = Y - \Delta s.
\eeq
The relation (\ref{Cl_eq}) is a generalization of the classical thermodynamic Clausius inequality $Q \geq -\Delta S$ to transitions between nonequilibrium steady states fitting in the phenomenological framework of steady state thermodynamics \cite{oono98, shib00}.  
The fluctuation theorem
\beq
\mean{e^{-\beta \Qhk}} = 1
\eeq
for the housekeeping heat has recently been proven for stochastic processes ruled by a Fokker-Planck equation \cite {spec05a}. It can be shown that this relation holds also for chemical master equations \cite{schm_dipl}.  

\section{An example}
\label{sec6}

\begin{figure}
 \includegraphics[width = 0.5\linewidth]{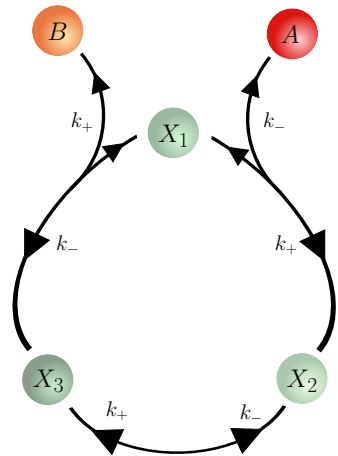} 
 \caption{Model reaction network (\ref{model_netw}) with three freely evolving species $X_1, X_2, X_3$ and two chemiostat species $A$ and $B$.}
\label{parad}
\end{figure}

We illustrate our general results by considering the simple model reaction network (see Fig. \ref{parad})
\begin{eqnarray}
A + X_1  \mathop{\rightleftharpoons}^{k_+}_{k_-}  X_2 ~~~
X_2 \mathop{\rightleftharpoons}^{k_+}_{k_-} X_3 ~~~
X_3 \mathop{\rightleftharpoons}^{k_+}_{k_-} X_1 + B
\label{model_netw}
\end{eqnarray} 
where $A$ and $B$ are chemiostat species (corresponding to $A_1$ and $A_2$ in our previous notation). For simplicity, we assume that all normalizing volumes $\dbv_j$ ($j = 1,2,3$) in principle defined through a microscopic Hamiltonian, see App.  \ref{App2}, are equal. These normalizing volumes connect the bare rates $k_{+,-}$ to the transition rates $w_{+,-}^{\rho}(\mathbf{n}, \ca)$ through (\ref{w+}) and (\ref{w-}). The chemiostat species have chemical potentials (\ref{def_mu_al})
\beq
\mu_A = E_A + T \ln \left( c_A \dbv_A \right)
\eeq
and
\beq
\mu_B = E_B + T \ln \left( c_B \dbv_B \right),
\eeq
where $E_A$ and $E_B$ are single particle energies of $A$ and $B$ species, respectively, and $\dbv_A, \dbv_B$ are appropriate normalizing volumes.
The stochiometric matrix $\mathbf{V} = (v_j^\rho)$ 
\beq
\mathbf{V} = 
\mathop {\left ( \begin {array}{r r r}
									-1&1&0\\
									0&-1&1\\
									1&0&-1\\
			\end{array}
		\right )}^{X_1\hspace{0.3cm}X_2~~X_3}
~~\begin{array}{c}{(\rho = 1)}\\{(\rho = 2)}\\{(\rho = 3)} \end{array}
\eeq		
gives the number of transformed $X_j$ molecules in one forward reaction of type $\rho$. Obviously, it is singular with $rank~ \mathbf{V} =  2$, which corresponds to case II from Sect. \ref{cases}. We therefore have the constraint 
\beq
n_1 + n_2 + n_3 = N
\eeq
on the accessible states of the phase space. Thus, the total number $N$ of $X$-molecules is a conserved quantity. 
\subsection{Equilibrium distribution}

In order to determine the potential equilibrium distribution, we try to solve eqs. (\ref{epsilonj})
\beq
 \left ( \begin {array}{r r r}
									-1&1&0\\
									0&-1&1\\
									1&0&-1\\
			\end{array}
		\right ) \cdot \left ( \begin {array}{c}
									\epsilon_1\\
									\epsilon_2\\
									\epsilon_3\\
			\end{array}
		\right ) =
			 		 \left ( \begin {array}{c}
									- \ln (c_A \dbv_A) + \ln(k_-/k_+)\\
									\ln(k_-/k_+)\\
									\ln (c_B \dbv_B) + \ln(k_-/k_+)\\
			\end{array}
		\right ),
\eeq						
which have solutions only if 
\beq
c_A / c_B = \left( \dbv_B / \dbv_A \right) (k_- / k_+) ^3 \equiv c_A^{\eq} / c_B^{\eq}.
\eeq
This is the expected condition for equilibrium since for any network cycle obeying detailed balance the product of forward and backward rates must be equal. We then get the solution
\beq
\left ( \begin {array}{c}
									\epsilon_1\\
									\epsilon_2\\
									\epsilon_3\\
			\end{array} \right )
 = 
		\left ( \begin {array}{c}
									\ln (c_A \dbv_A) - \ln (k_-/k_+)\\
									0\\
									\ln(k_-/k_+)\\
			\end{array} \right)
+
\gamma \cdot \left ( \begin {array}{c}
									1\\
									1\\
									1\\
			\end{array} \right),
\eeq
with arbitrary $\gamma$. The equilibrium distribution is the projection of the Poisson distribution (\ref{eq_distr}) to the accessible state space, i. e. the states with total number $N$ of $X$-molecules. This yields
\begin{eqnarray}
p^{\eq}(\mathbf{n}, \ca) &\propto& \prod_j \frac {\Om^{n_j}} {n_j!} e^{-\epsilon_j n_j} \delta_{N,n_1+n_2+n_3} \nn \\ 
&\propto& \frac {\delta_{N,n_1+n_2+n_3}} {n_1! n_2! n_3!}  \left(\frac {k_-}{k_+ c_A \dbv_A}\right)^{n_1} \left(\frac {k_+}{k_-}\right)^{n_3} .
\end{eqnarray}

\subsection{Energy}

In principle, the energies $E_j$ of one molecule $X_j$ are defined as the thermodynamic average of the microscopic Hamiltonian, see App. \ref{App2}.
Without specifying such a Hamiltonian, for given bare rates $k_{+,-}$, the energies $E_j$ of a single solute molecule $X_j$ must obey (\ref{def_E_j}) 
\beq
\left ( \begin {array}{r r r}
									-1&1&0\\
									0&-1&1\\
									1&0&-1\\
			\end{array}
		\right ) \cdot \left ( \begin {array}{c}
									E_1\\
									E_2\\
									E_3\\
			\end{array}
		\right ) = \left ( \begin {array}{r}
									\ln (k_-/k_+)\\
									\ln(k_-/k_+)\\
									\ln(k_-/k_+)\\	
			\end{array}
		\right ) - \left ( \begin {array}{c}
									-E_A\\
									0\\
									E_B\\
			\end{array}
		\right )
\eeq
which has a solution for the $E_j$ only if
\beq
E_B - E_A = 3 \ln \frac {k_-}{k_+}.
\label{E_al_bsp}
\eeq
This is the condition for the reaction network to be thermodynamically consistent as discussed in Sect. \ref{rel}. To simplify matters, we have set here and in the following the inverse temperature $\beta = 1$.

The difference in energies $E_A$ and $E_B$, combined with entropic forces, is the driving force of a potential nonequilibrium steady state. If the energies were equal, there was no reason for the reaction to run on average in one direction for equilibrated concentrations $c_A \dbv_A = c_B \dbv_B$ and hence the bare rates $k_+$ and $k_-$ were equal. The ratio of bare rates $k_- / k_+$ thus depends on the reservoir energies through (\ref{E_al_bsp}).
We then get for the relationship between energies $E_j$ and bare rates $k_{+,-}$ 
\beq
\left ( \begin {array}{c}
									E_1\\
									E_2\\
									E_3\\
			\end{array}
		\right ) = \left ( \begin {array}{c}
									E_0 - E_A\\
									E_0 + \ln(k_-/k_+)\\
									E_0 + 2\ln(k_-/k_+)\\	
			\end{array}
		\right )
\label{E_ex}
 \eeq
with arbitrary $E_0$. 

Having obtained single particle energies in an equilibrium situation, the reaction system can now be driven out of equilibrium by changing the concentration of A or B molecules. Non-adiabatic transitions between equilibrium distributions are achieved by driving $c_A(\tau)= \left( \dbv_B / \dbv_A \right) (k_- / k_+) ^3 c_B(\tau)$. Nonequilibrium steady states and transitions between them are obtained for any protocol with $c_A(\tau) \neq \left ( \dbv_B / \dbv_A \right) (k_- / k_+) ^3  c_B(\tau)$.

\subsection{First law}

Whenever a single reaction takes place, the chemiostats associated with $A$ and $B$ molecules apply work to the system consisting of the $X_j$-molecules. Furthermore, the system of the $X_j$-molecules dissipates heat into the surrounding aequous solution acting as a heat bath.
The chemical work $\Wc$ done by the chemiostats due to a single reaction is generally given by (\ref{W_single}) as 
\beq
\Wc^{r, \rho} = -\sum_\al \mu_\al \Delta n_\al^{r, \rho}.
\eeq
Due to energy conservation, the dissipated heat for one reaction step is given by (\ref{Q_single}) as 
\beq
Q^{r, \rho}  = \Wc^{r, \rho} - \sum_{j} E_{j} \Delta n_j^{r, \rho}.
\eeq
Having defined work, dissipated heat and internal energy, we can formulate the first law in the form (\ref{first_law})
\beq
\Wc^{r, \rho} = \Delta E^{r, \rho} +  Q^{r, \rho}.
\label{first_law1}
\eeq

Chemical work, dissipated heat and change of internal energy for one single forward reaction step of each type $(\rho = 1,2,3)$ are shown in Tab. \ref{tab1}. 
\begin{table}
\begin{tabular}{c||c|c|c}
\hline
&$\mathbf{ A + X_1  \mathop{\rightleftharpoons}^{k_+}_{k_-}  X_2}$&$\mathbf{X_2 \mathop{\rightleftharpoons}^{k_+}_{k_-} X_3}$&$\mathbf{X_3 \mathop{\rightleftharpoons}^{k_+}_{k_-} X_1 + B}$\\
\hline
\hline
$\Wc$& $\mu_A $&$ 0 $&$ -\mu_B$ \\
\hline
$Q$& -$E_2 + E_1 + \mu_A $&$ - E_3 + E_2 $&$ -E_1 + E_3 - \mu_B $\\
&$ = \ln \left ( \frac {k_+ c_A \dbv_A}{k_-} \right )$&$ = \ln \left ( \frac {k_+}{k_-} \right )$&$ = \ln \left ( \frac {k_+ }{k_-c_B \dbv_B} \right )$\\
\hline
$\Delta E $&$\ln(k_-/k_+) + E_A$&$\ln(k_-/k_+)$&$-E_A - 2 \ln(k_-/k_+)$\\
\hline
\end{tabular}
\caption{Work, dissipated heat and change of internal energy for a single step in forward direction for the reaction network (\ref{model_netw}).}
\label{tab1}
\end{table} 
The expressions for the heat agree with the general result (\ref{Q_rates}). The dissipated heat for a given sequence of reactions thus depends on the concentrations $c_A$ and $c_B$. In equilibrium, the dissipated heat is zero for a complete cycle and therefore is zero on average. However, for a single reaction step the heat is not necessarily zero.

\subsection{Entropy production}

For an illustration of the entropy change, we calculate the total entropy production (\ref{dstot}) due to one occuring forward reaction 
\beq
X_3 \mathop{\rightleftharpoons}^{k_+}_{k_-} X_1 + B.
\eeq
from state $\n^- \equiv (n_1, n_2, n_3)$ to state $\n^+ \equiv (n_1+1, n_2, n_3-1)$ at time $\tau$.
The total entropy production $\Delta \stot$ (\ref{dstot}) is the sum of the change in system entropy (\ref{system_entropy})
\begin {eqnarray}
\Delta s &=& -\ln p(\mathbf{n^+}, \tau) + s^0(\mathbf{n^+}) + \ln p(\mathbf{n^-}, \tau) - s^0(\mathbf{n^-}) \nn \\
&=& \ln \frac { p(\n^-, \tau) n_3} {p(\n^+, \tau) (n_1 + 1)}
\end {eqnarray}
and the entropy change of the heat bath (\ref{Q_rates})
\beq
\Delta \sm = Q = \ln \left ( \frac {k_+ }{k_-c_B \dbv_B} \right ).
\eeq
The total entropy production then is
\begin{eqnarray}
\Delta \stot &=& \Delta s + \Delta \sm  \nn \\
&=& \ln \frac {p(\n^-, \tau)} {p(\n^+, \tau)} + \ln \frac {n_3  k_+} {(n_1+1) k_- c_B \dbv_B} \nn \\
&=& \ln \frac {p(\mathbf{n^-, \tau}) w_+^{\rho} (\mathbf{n^-}, \ca(\tau))}{p(\mathbf{n^+}, \tau) w_-^{\rho} (\mathbf{n^+}, \ca(\tau))} = R.
\end {eqnarray}
This illustrates the cancellation of the degeneracy terms and the equivalence of the abstract quantity $R$ (\ref{R_allg}) and the thermodynamically motivated entropy production $\Delta \stot$ (\ref{dstot}). The total entropy production thus could be extracted from experiments if the state probabilities and the transition probabilities of the stationary state could be measured.

\section{Conclusion and perspectives}

In summary, for arbitrary chemical networks in a dilute solution described by a chemical master equation, we have developed a consistent thermodynamic description along a single stochastic trajectory by identifying work, internal energy and dissipated heat for each reaction step. A unique identification of internal energy and dissipated heat requires a microscopic Hamiltonian for the interaction between the reacting species and the solvent and cannot be extracted from the stochastic dynamics alone. However, we have shown that work and total entropy production do not depend on the microscopic details and can be extracted from the dynamics once the transition rates in the chemical master equation are specified or measured in an experiment. Entropy production involves both the change of the entropy of the medium given by the dissipated heat and genuine entropy of the network which requires a proper consideration of a degeneracy factor. The total entropy production fulfills an integral fluctuation theorem for arbitrary driving and a detailed fluctuation theorem in the steady state. 

This systematic approach relies on three crucial concepts. 
First, an energetic interpretation of stochastic master equation dynamics requires the comparison with an appropriate equilibrium state whose microscopic Hamiltonian for the interaction with the solvent must be known. By choosing the equilibrium concentrations of reservoir particles and thus adjusting the net flux to zero, a stationary distribution obeying detailed balance can be obtained for any thermodynamically consistent reaction network. Second, the persistence of mass action law kinetics even in non-equilibrium
leads to the concept that energy differences are related to the ratio of bare rates in the same way as in equilibrium. Crucial for the persistence of this assumption is the time scale separation of diffusion and reaction time constants. For higher concentrations or long range interactions, a coupled system of reaction-diffusion equations would be necessary. The general concepts of stochastic thermodynamics would still apply; however, the expressions for energy, dissipated heat, work and entropy would get more complex.
Third, we have decided to define the system as the collection of those species whose numbers we can follow in principle. The internal energy then is associated with these molecules only.
This third assumption could be somewhat relaxed since one could always include one of the chemiostat species to the system species. The first two assumptions, however, are crucial if one wants to keep the energetic interpretation. Of course it is still possible to derive integral and detailed fluctuation theorems for a stochastic entropy production in any network obeying master equation dynamics \cite{seif05a}. The interpretation of the abstract quantity $\Delta \sm$ as an exchanged heat, however, requires the first law energy balance and the correct identification of the degeneracy factor.

Distributions of work or entropy production may be used to reconstruct thermodynamic equilibrium quantities or to characterize the network, similar to the reconstruction of free energy landscapes for unzipping or unfolding transitions using the Jarzynski relation \cite{coll05, brau04, humm01}. A first step in this direction is the use of fluctuation theorems to determine chemical driving forces in reaction cycles \cite{min05}.
Of particular interest would also be the derivation of the distributions of work and total entropy production for frequently used models of gene regulation networks or signal processing networks, especially since it is known from a simple athermal model system that these distributions can show a quite rich structure \cite{schu05, tiet06}.

While there exists already a large amount of experimental work on fluctuation theorems for Langevin-systems, we are not aware of any attempts to probe fluctuation theorems for chemical reaction networks experimentally. Such experiments could contribute substantially to the ongoing effort to achieve a better understanding of nonequilibrium systems.

\begin{acknowledgements}
%\section*{Acknowledgement}
We thank C. Jarzynski and T. Speck for inspiring discussions.
\end{acknowledgements}

%%%%%%%%%%%%%%%%%%%%%%%%%%%%%%%%%%%%%%%%%%%%%%%%%%%%%%%%%%%%%%%%%%%%%%%
%%%%%%%%              Appendix                %%%%%%%%%%%%%%%%%%%%%%%%%
%%%%%%%%%%%%%%%%%%%%%%%%%%%%%%%%%%%%%%%%%%%%%%%%%%%%%%%%%%%%%%%%%%%%%%%

\appendix
\vskip 1cm

\section{Normalizing volumes and Thermodynamics of a dilute solution}
For a precise definition of the normalizing volumes $\dbv_\al, \dbv_j$ and the energies $E_\al$, $E_j$, we will now briefly review the standard thermodynamics of a dilute solution as found in textbooks, see e. g. \cite{schwabl}. For ease of notation, we treat the case of a dilute solution of $N'$ equivalent solute molecules in a solvent with $N$ molecules. This calculation can easily be generalized to the case of more than one species and applies both to the $A_\alpha$ molecules and to the $X_j$ molecules. 
We assume that the number of solvent molecules $N$ is fixed and therefore use a semi-grandcanonical ensemble. In the dilute limit $N' \ll N$, the Hamiltonian can then be split into the contributions
\begin{eqnarray}
H^{N,N'} (\{\mathbf{r}\}, \{\mathbf{r'}\}) &=&  H^{N}_1(\{\mathbf{r}\}) + H^{N'}_1(\{\mathbf{r'}\}) \nn \\
&&+ H^{N,N'}_{I}(\{\mathbf{r}\}, \{\mathbf{r'}\})
\end{eqnarray}
with
\beq
H^{N'}_1(\{\mathbf{r'}\}) \equiv \sum_{i=1}^{N'} H_1(\mathbf{r_i'})
\eeq
and
\beq
H^{N,N'}_{I}(\{\mathbf{r'}\}, \{\mathbf{r}\}) \equiv  \sum_{i=1}^{N'} H_{I}(\mathbf{r_i'}, \{\mathbf{r}\}),
\eeq
where $\mathbf{r_i'}$ denotes the position of the $i$th solute particle with $i = 1,\dots,N'$ and $\{ \mathbf{r} \}$ are the coordinates of the solvent molecules.
One particle energies $E_i$ ($i = \al, j$) are defined through
\beq
E_i = \mean{H_1(\mathbf{r_i'}) + H_1(\mathbf{r_i'}, \lbrace \mathbf{r} \rbrace)}_{N, N' = 1} . 
\label{E_s}
\eeq
Here, $\mean{.}$ denotes the canonical average with the appropriate canonical Boltzmann-weight.

Having defined one particle energies $E_\al$ from a microscopic point of view through (\ref{E_s}), one can transform the chemical potential (\ref{def_mu_al_kon}) into the form (\ref{def_mu_al}) by choosing an appropriate normalizing volume 
\beq
\omega_\alpha \equiv e^{\beta (\mu^0_\al - E_\al)} / c_0.
\label{omeg_al}
\eeq
where $\beta \equiv 1 /  T$ is the inverse temperature. 

We now recall how to obtain the grandcanonical equilibrium distribution $p_N^{eq}(N')$. Integrating out the spatial degrees of freedom, we get
\beq
p_N^{eq}(N') = \frac {Z_{can}^{N, N'} e^{\beta \mu N'}}{\sum_{N'} Z_{can}^{N, N'} e^{\beta \mu N'}}
\label{p_N}
\eeq
where $Z_{can}^{N, N'}$ is the canonical partition function which can be calculated as %in the limit $N'/N \to 0$ as
\beq
Z_{can}^{N, N'} = F(N) \frac 1 {N' !} \left ( \frac \Omega {\lambda'^3} \Psi(T, \Omega /N) \right)^{N'}
\label{Z_can}
\eeq
where $\Psi(T, \Omega /N)$ and $F(N)$ depend on the particular Hamiltonian and $\lambda'$ is the thermal de Broglie wavelength of a solute particle \cite{schwabl}.
The solute molecules thus behave like an ideal gas in the effective (constant) potential $\Psi(T, \Omega/N) / \beta$. 
Inserting (\ref{Z_can}) into (\ref{p_N}) yields
\begin{eqnarray}
p_N^{eq}(N') &\propto& \frac 1 {N'!} \left ( \frac \Omega {\lambda'^3} \right)^{N'} \Psi^{N'} e^{\beta \mu N'} \nn \\
 &=& \frac 1 {N'!} \left ( \frac {\Omega \Psi} {\lambda'^3 e^{-\beta E}}\right)^{N'} e^{-\beta(E-\mu)N'}
\label{p_N1}
\end{eqnarray}
Thus, Eq. (\ref{p_N1}) has the form of eq. (\ref{grandcan_distr}) with 
\beq
\omega \equiv {\lambda'^3 e^{-\beta E}} / {\Psi}.
\label{omeg_j}
\eeq
A simple calculation shows that eqs. (\ref{def_mu_al}) and (\ref{omeg_al}) are consistent with eq. (\ref{omeg_j}).
\label{App2}

\section{Analysis of case II from Sect. \ref{cases}}

If there exists more than one solution of (\ref{epsilonj}), the rank of the stochiometric matrix $\mathbf{V}$ is smaller than the number $\Nj$ of species $X_j$. This is usually the case if there are less reactions $\rho$ than species $X_j$, i.e. $\Nrho < \Nj$. We then have $\Nmu = \Nj - rank \mathbf{V}$ (generically : $\Nmu = \Nj - \Nrho$) constraints 
\beq
\sum_j K^{\mu}_j n_j = L^{\mu}
\eeq
$\mu = (1,\dots,\Nmu)$ for the accessible states with $\Lv = (L^1,\dots,L^{\Nmu})$ depending on the initial distribution. The matrix $\mathbf{K}$ has rows which span the nullspace of the stochiometric matrix $\mathbf{V}$
\beq
\sum_{j=1}^{\Nj} (v_j^{\rho}) K^{\mu}_j = \mathbf{0}.
\eeq
The general solution of (\ref{epsilonj}) then is the sum of a special solution $\epsilon_j^0$ and a nullspace vector $\mathbf{K x}$
\beq
\epsilon_j = \epsilon_j^0 + \sum_{\mu = 1}^{\Nmu} K_j^{\mu} x^{\mu}
\eeq
for an arbitrary vector $\mathbf{x} \in \mathbf{R}^{\mu}$.
The state space $\{\mathbf{n}\}$ is then separated into an infinite number of subspaces $\chi_{\Lv} = \{\mathbf{n}| K^{\mu}_j n_j = L^{\mu} \} \subset \{\mathbf{n}\}$. If all possible initial states lie in the same subspace $\chi_{\Lv}$, the equilibrium distribution is the projection of the Poissonian (\ref{eq_distr1}) to the subspace $\chi_{\Lv}$ with the $\epsilon_j$ given by any solution of (\ref{epsilonj}). It does not matter which solution of (\ref{epsilonj}) is chosen since

\begin{eqnarray}
p^{\eq}(\mathbf{n}, \ca) &=& \mathcal N \prod_j \frac {\Om^{n_j}} {n_j!} e^{-\epsilon'_j n_j} \nn \\&=& 
\mathcal N \prod_j \frac {\Om^{n_j}} {n_j!} e^{-\left(\epsilon_j + \sum_{\mu} K_j^{\mu} x^{\mu} \right) n_j} \nn \\
&=& \mathcal N \prod_j \frac {\Om^{n_j}} {n_j!} e^{-\left(\epsilon_j n_j\right) - L_j}  \nn \\ &=&  \mathcal N' \cdot \prod_j \frac {\Om^{n_j}} {n_j!} e^{-\epsilon_j n_j}.
\end{eqnarray}
Projections comprise normalization and therefore we get the same distribution for different solutions of (\ref{epsilonj}).

\label{App1}

\end{document}